\documentstyle[12pt]{article}

\newcommand{\EQ}{\begin{equation}}
\newcommand{\EN}{\end{equation}}
\begin{document}
\topmargin 0pt
\oddsidemargin=-0.4truecm
\evensidemargin=-0.4truecm
\renewcommand{\thefootnote}{\fnsymbol{footnote}}
\newpage
\setcounter{page}{0}
\begin{titlepage}
\begin{flushright}
IC/92/196,~~SISSA-140/92/EP,~~LMU-09/92\\
August 1992
\end{flushright}
\vspace*{-0.2cm}
\begin{center}
{\large  PLANCK SCALE EFFECTS IN NEUTRINO PHYSICS}
\vspace{0.4cm}

{\large Eugeni Kh. Akhmedov${}^{(a,b,c)}$
\footnote{E-mail: akhmedov@tsmi19.sissa.it, ~akhm@jbivn.kiae.su},
Zurab G. Berezhiani${}^{(d,e)}$
\footnote{Alexander von Humboldt Fellow}
\footnote{E-mail: zurab@hep.physik.uni-muenchen.de,
{}~vaxfe::berezhiani},\\
Goran Senjanovi\'{c}${}^{(a)}$
\footnote{E-mail: goran@itsictp.bitnet, ~vxicp1::gorans},
Zhijian Tao${}^{(a)}$}\\
\vspace*{0.6truecm}
${}^{(a)}$\em{International Centre for Theoretical Physics,
I-34100 Trieste, Italy\\
${}^{(b)}$Scuola Internazionale Superiore di Studi Avanzati,
I-34014 Trieste, Italy\\
${}^{(c)}$Kurchatov Institute of Atomic Energy,
Moscow 123182, Russia\\
${}^{(d)}$Sektion Physik der Universit\"{a}t M\"{u}nchen,
D-8000 Munich-2,
Germany\\
${}^{(e)}$Institute of Physics, Georgian Academy of Sciences, Tbilisi
380077, Georgia\\}
\end{center}
\begin{abstract}
We study the phenomenology and cosmology of the Majoron (flavon) models
of three active and one inert neutrino paying special attention to the
possible (almost) conserved generalization of the
Zeldovich-Konopinski-Mahmoud lepton charge. Using Planck scale physics effects
which provide the breaking of the lepton charge, we show how in this picture
one can incorporate the solutions to some of the central issues in neutrino
physics such as the solar and atmospheric neutrino puzzles, dark matter and
a 17 keV neutrino. These gravitational effects induce tiny Majorana mass terms
for neutrinos and considerable masses for flavons. The cosmological demand for
the sufficiently fast decay of flavons implies a lower limit on the electron
neutrino mass in the range of 0.1-1 eV.
\end{abstract}
\vspace{0.5cm}
\end{titlepage}
\renewcommand{\thefootnote}{\arabic{footnote}}
\setcounter{footnote}{0}
\newpage
\section{Introduction}
The central open issues in neutrino physics, according to our belief,
are

{}~{\bf (a) The solar neutrino puzzle (SNP)}.
The solar neutrino experiments under operation \cite{SNE1,SNE2,SNE3,SNE4}
indicate a deficiency of solar neutrinos pointing to neutrino properties
being a source of the discrepancy between theory and experiment. The most
popular and natural explanation is based on oscillations of $\nu_e$ into
another neutrino in solar matter or in vacuum during the flight to the earth.

{}~{\bf (b) The atmospheric neutrino puzzle (ANP)}.
There is some evidence for a significant depletion of the atmospheric
$\nu_{\mu}$ flux, by almost a factor of two \cite{ANP}. This result, if
true, would point again to neutrino oscillations, this time of $\nu_{\mu}$
into another species, with a large mixing angle and an oscillation length
less than or of the order of the atmospheric height.

It is, at least in principle, possible to resolve both the SNP and ANP in
the context of the usual three neutrino flavors, e.g. the SNP could be
due to the $\nu_{e}\rightarrow \nu_{\mu}$ oscillations, and the ANP
due to the $\nu_{\mu}\rightarrow \nu_{\tau}$ oscillations.

{}~{\bf (c) Dark matter problem}.
Neutrinos with a mass in the range of 10--100 eV have been considered for
many years as natural candidates for dark matter needed to explain the
observed large scale structure of the universe. This popular, so-called
hot dark matter (HDM) scenario, also able to explain the missing
cosmological density, was disfavored in the last years due to the bounds
on the primordial density fluctuations coming from the measurements of the
cosmic microwave background radiation (CMBR). The recent COBE discovery of
the CMBR anisotropy \cite{COBE}, however, suggests at least some presence of
HDM together with cold dark matter (CDM) with the latter being the dominant
component \cite{Shafi}. This role can now be naturally played by neutrinos
with a mass in eV range.

{}~{\bf (d) A 17 keV neutrino}.
As exciting as it is, the existence of the 17 keV neutrino is far from being
established \cite{PRO,CONTRA}. Many theoretical models on the subject were
proposed \cite{THEOR}, but the difficult task of incorporating the SNP in this
picture has only recently been addressed \cite{X}. The problem is that the
conventional scenario of three neutrinos $\nu_e$, $\nu_{\mu}$ and $\nu_{\tau}$
cannot reconcile laboratory constraints with solar neutrino deficit. Namely,
the combined restriction from the neutrinoless double $\beta$ decay and
$\nu_e\leftrightarrow \nu_{\mu}$ oscillations leads to a conserved (or at most
very weakly broken) generalization of ZKM \cite{ZKM} symmetry:
$L_e-L_{\mu}+L_{\tau}$ \cite{V}. This in turn implies
the 17 keV neutrino $\nu_{17}$ mainly to consist of $\nu_{\mu}^{c}$ and
$\nu_{\tau}$, mixed by the Simpson angle $\theta_S\simeq 0.1$ with the
massless $\nu_e$. Clearly, in this picture there is no room for the solution
of the SNP due to neutrino properties.

It is well-known by now that the LEP limit \cite{LEP} on $Z^0$ decay width
excludes the existence of yet another light active neutrino. However, the same
in general is not true for a sterile neutrino ($n_R$). Of course, once
introduced, $n$ (instead of $\nu_{\mu}^{c}$) can combine with the $\nu_{\tau}$
to form $\nu_{17}$ or just provide a missing light partner to $\nu_e$ needed
for the neutrino oscillation solution to the SNP. The latter possibility has
been recently advocated by the authors of ref. \cite{X}. In this paper we
study in some detail the physics of an extra sterile neutrino. We will show
that its existence can accommodate the solution to all the above puzzles. We
offer a systematic study of this scenario, paying special attention to
possible  effective operators that could induce neutrino masses. We consider
the case of a maximal abelian lepton flavor symmetry with $n_R$ included,
inspired by an analysis performed by Barbieri and Hall (BH) \cite{BH} for the
case of three active neutrinos. The crucial characteristics of this approach
is the existence of flavons, i.e. Majorons associated with spontaneous
violation of extended lepton flavor symmetries. These flavons can naturally
provide sufficiently fast decay of $\nu_{17}$ which is necessary for
cosmological reasons.

In order to generate neutrino oscillations in the light sector needed for the
solution of the SNP and ANP, it will turn out necessary to break the lepton
number symmetry. We propose an interesting possibility of higher dimensional
operators being responsible for this breaking \cite{HDO,GLR}. These operators
could naturally result from the quantum gravitational effects and should be
cut off by the Planck scale. We find it encouraging that such tiny effects may
be sufficient for the simultaneous solution of the above mentioned problems.
It will be shown in section 3 that these effects induce mass splittings
between the components of
Dirac or ZKM neutrinos of the order of $10^{-6}$ eV. Since the solution to ANP
seems to require $\Delta m^2\simeq 10^{-2}-10^{-3}$ eV$^2$ with large mixing
angles, this in turn suggests that the mass of the heavy neutrino is of the
order of a few keV. This encouraged us to seriously pursue the possibility of
a controversial 17 keV neutrino, although all we need is the existence of a
heavy neutrino with a mixing angle which could be much smaller than
$\theta_S$. We would like to emphasize that otherwise our analysis is quite
general, and it will hold true even if $\nu_{17}$ with $\theta_S \simeq 0.1$
disappears\footnote{Hereafter $"\nu_{17}"$ denotes a heavy neutrino, even if
its mass is not exactly 17 keV and its mixing is different from the Simpson
angle.}.
Furthermore, the same gravitational effects create the potential problem by
inducing appreciable masses for flavons, of the order of 1 keV. Just like
$\nu_{17}$, they also must decay fast enough in order not to postpone the
matter dominated era of the expansion of the universe needed for the
development of the cosmological large scale structure. This requirement is put
on the firmer ground through the COBE findings indicating rather small initial
density fluctuations. Since the couplings of Majorons to neutrinos are
necessarily proportional to the masses of the latter, this leads to both
phenomenologically and cosmologically important lower limit on the electron
neutrino mass $m_{\nu_e}>(0.1-1)$ eV. As we will show in the section 3,
electron neutrino thus becomes a natural candidate to provide the needed
10--30 per cent hot dark matter of the universe.

Our paper is organized as follows. In the next section we offer a general
study of a system of three active and one sterile neutrino with a conserved
generalized ZKM lepton number. In the section 3 we study the implications of
the necessary breaking of this symmetry induced through the quantum
gravitational effects. In section 4 a specific model is offered for the sake
of demonstration and finally the last section is reserved for the discussion
and outlook.

\section{The effective operator study of the neutrino masses}
The introduction of a new state $n_R$ opens up a number of new possibilities
for a conserved (or approximately conserved) generalized lepton number $L$.
We distinguish two such different classes.

(i) A case of one Dirac and one ZKM state, for which $L$ takes the form
\begin{equation}
L_{\pm}=L_e-L_{\mu}\pm(L_{\tau}-L_{n^{c}})
\end{equation}
where hereafter we use the convenient notation of a left-handed $n^{c}$
field $(n^{c})_{L}\equiv C{\bar n}_{R}^{T}$. Notice that $L_e$ and
$L_{\mu}$ charges must enter with opposite signs in order to comply with
the absence of $\nu_e\leftrightarrow \nu_{\mu}$ oscillation with a
Simpson mixing angle $\theta_S\simeq 0.1$. In each of $L_+$ and $L_-$ case,
we are still left with the freedom of having $\nu_{17}\simeq\nu_{\tau}+n_R$ or
$\nu_{17}\simeq\nu_{\tau}+\nu_{\mu}^c$ for $L_{+}$ and
$\nu_{17}\simeq\nu_{\mu}+n_{R}$ or $\nu_{17}\simeq \nu_{\tau}+n_R$ for
$L_{-}$. If $n_{R}$ is a part of $\nu_{17}$, one is potentially in conflict
with the supernova 1987A bound $m_{17}\leq (1-30)$ keV due to $n$ being
sterile and taking away the energy of the supernova after a
helicity flipping scattering $\nu_{\tau}(\nu_{\mu})\rightarrow n$ \cite{SNB}.
The same limit does not apply, of course, when $\nu_{17}$ consists of only
active neutrinos.

(ii) A case of either Dirac or ZKM $\nu_{17}$ and two massless states,
corresponding to lepton charges with only one minus sign:
\begin{eqnarray}
L_1=L_e-L_{\mu}+L_{\tau}+L_{n^{c}}\nonumber \\
L_2=L_e+L_{\mu}-L_{\tau}+L_{n^{c}}\\
L_3=L_e+L_{\mu}+L_{\tau}-L_{n^{c}}\nonumber
\end{eqnarray}
Obviously, ${}-L_e$ is not allowed.

In what follows, we shall analyse systematically the above possibilities, some
of which were studied in the context of specific models \cite{SpM}. Our aim
is to extract as much as possible model independent information, but we will
also present a simple model which will serve as an illustration of general
considerations.

One may wonder at this point whether $\nu_{i}\rightarrow n$ oscillations
could bring $n$ into equilibrium before the nucleosynthesis
\cite{CL1,CL2,CL3}, thereby affecting the primordial ${}^4$He abundance in
the universe \cite{Chic}. The situation crucially depends on the mixing angle
$\theta_n$ between sterile and active neutrinos and so varies with the
structure of $\nu_{17}$.

The relative presence of $n$ in the number of neutrino species at the time of
nucleosynthesis is of course a function of its decoupling temperature $T_n$.
We can thus speak of two distinct cases, $T_n>T_{QCD}$ and $T_n\leq T_{QCD}$,
where $T_{QCD}$ is the QCD phase transition temperature.
In the former case, it can be shown that $n$ counts at most $0.3$ of the
usual neutrino contribution due to the reheating of active neutrinos when
$T$ drops below $T_{QCD}$, whereas in the latter case we expect $N_{\nu}
\simeq 4$, since the only change below $T_{QCD}$ is the annihilation of
$\mu^+\mu^-$ pairs, which barely changes the temperature of the neutrino
sea. Clearly, $T_n$ depends on the mixing angle $\theta_n$, the smaller
$\theta_n$ is, the larger $T_n$.

{}From an analysis of ref. \cite{CL1} one can get (for a mass difference
$\Delta m\simeq 17$ keV) the relation between $\theta_n$ and $T_n$:
\vspace{.2truecm}
\EQ
T_{n}^{3} \simeq \frac{(3~{\rm MeV})^3}{1/2\sin^{2}2\theta_{n}}
\EN
\vglue .1truecm
\noindent
and $T_n\ge T_{QCD}\simeq 200$ MeV requires $\theta_n\le 10^{-3}$.

Now it is readily seen that for $L_{-}$ the mixing angle $\theta_n$ coincides
with $\theta_S$ and therefore in this case one predicts $N_{\nu}\simeq4$
(since $T_n\ll T_{QCD}$). In other cases the situation depends on the details
of the model, i.e. on the structure of $\nu_{17}$; we will return to them
later when we discuss the neutrino mass matrix.

Before proceeding, we wish to recall the fact that $\nu_{17}$ must decay
fast enough in order to comply with cosmological constraints, and it
appears that the simplest mechanism is provided by the Majoron, the Goldstone
boson of a spontaneously broken lepton number (or lepton flavor) symmetry.
We therefore assume large global symmetry $G$ spontaneously broken down
to $L$. In particular, this implies the existence of some new scalar
fields, generically denoted by $S$, which transform nontrivially under $G$.
Since the Majoron (one or more) is a phase of $S$, due to already mentioned
LEP constraints on the $Z^0$ decay width $S$ fields should be singlets
under $SU(2)_L\times U(1)$. Furthermore, any effective mass term invariant
under $G$ will necessarily involve $ S$ fields (assuming that the lepton
flavor numbers, including $n$, are distinct). When an illustration is needed,
we discuss the straightforward extension of lepton number based on
$G=U(1)_e\times U(1)_{\mu}\times U(1)_{\tau}\times U(1)_{n}$.

In order to generate small masses naturally, we allow no tree-level $d=4$
operators that could lead to neutrino masses. In particular, this implies that

\noindent
{}~~(a) the only scalar fields are $SU(2)_L$ doublets and singlets, and\\
\mbox{~~}(b) no singlet carries such quantum numbers under $G$ as to
give direct $(d=4)$ Yukawa couplings.\\
In the context of the above example we allow only $S_{ab}$, $a\not=b$
($a,b=e,\mu,\tau, n$) singlet fields. These fields give naturally rise to
"flavons", i.e. Majorons which change lepton flavor and provide fast
decay of the $\nu_{17}$ \cite{BH}.

Consistent with the above rules, the leading effective Yukawa neutrino
operators invariant under $SU(2)_L\times U(1)\times G$ are
\vspace*{.2truecm}
\EQ
\alpha_{ij}(l_i^TC\tau_2\vec{\tau}l_j)\frac{H^T\tau_2\vec{\tau}H}{M^2}
S_{ij}^*,~~~
\alpha_{in}(l_{i}^{T}Cn^c)\frac{\tau_2 HS_{ij}^{*}S_{jn}}{M^2}
\EN
\vglue .1truecm
\noindent
where $l_{i}^T=(\nu_{iL}^T,~e_{iL}^T)$ are the leptonic weak doublets,
$H$ is the usual $SU(2)_L\times U(1)$ Higgs doublet, $M$ is a
regulator (cut-off) scale which is an input parameter and should be above
$<H>\sim M_W$ and $<S>$ ($S$ generically denotes $S_{ij}$ and $S_{in}$), and
$\alpha_{ab}$ are dimensionless factors expected to arise from the loop
expansion, $\alpha_{ab}\leq 10^{-3}-10^{-4}$. The quantum numbers of $S_{ab}$
fields under $G=U(1)_e\times U(1)_{\mu}\times U(1)_{\tau}\times U(1)_{n}$ are
\begin{eqnarray}
S_{e\mu}\hskip 0.2pc (1,1,0,0) \hskip 1pc S_{en}\hskip 0.2pc (1,0,0,
1)\nonumber \\
S_{\mu\tau}\hskip 0.2pc (0,1,1,0) \hskip 1pc S_{\mu n}\hskip 0.2pc (0,1,
0,1)\\
S_{e\tau}\hskip 0.2pc (1,0,1,0) \hskip 1pc S_{\tau n}\hskip 0.2pc (0,0,
1,1)\nonumber
\end{eqnarray}
Leptons carry their usual flavor charges, $n^{c}$ carries ${}-1$ unit of
$U(1)_{n}$, and the Higgs doublet $H$ of course carries no lepton flavor.
{}From the constraints on $\nu_{17}$ flavon decay, one can deduce the limit
30 GeV$\leq <S>\leq 300$ GeV \cite{BH}, where the non-vanishing $<S>$
conserve lepton number $L$ (for any $L$ defined above there corresponds
a certain set of $<S>$).

Before one specifies the form of $\nu_{17}$ in the sense discussed above,
one cannot decide the value of $M$ and $<S>$. For example, if $\nu_{17}
\simeq\nu_{\tau}+n$, both $<S>$ and $M$ could be as large as desired,
whereas in the case $\nu_{17}\simeq\nu_{\tau}+\nu_{\mu}^{c}$ both $M$ and
$<S>$ should be close to $M_W$ (see eq. (21) below). We come back to this
question in the specific examples, suffice is to say that the operators (4)
give the leading contributions to neutrino masses. We start for definiteness
with $L_+=L_e-L_{\mu}+L_{\tau}-L_{n^{c}}$, in which case the non-vanishing
VEVs are $<S_{e\mu}>$, $<S_{\mu\tau}>$ and $<S_{\mu n}>$. The neutrino mass
matrix in the Dirac basis takes then the form
\begin{equation}
\begin{array}{cc}
 & {\begin{array}{cc} n^{c} & \nu_{\mu}\end{array}}\\
M_{\nu}~=~\begin{array}{c}
\nu_e\\ \nu_{\tau}\end{array}&{\left(\begin{array}{cc}
m_{en}&m_{e\mu}\\m_{\tau n}&m_{\tau\mu}\end{array}\right)}\end{array}
\end{equation}
\noindent
{}From the smallness of the Simpson's angle and the absence of the $\nu_{\mu}
\rightarrow X$ oscillations it follows that only one entry of $M_{\nu}$,
either $m_{\tau n}$ or $m_{\tau\mu}$, can be $\sim$17 keV, whereas the other
entries must be at least an order of magnitude smaller. As we mentioned
before, there is still freedom for $\nu_{17}$ to consist of either
(a) $\nu_{\tau}+n$ or (b) $\nu_{\tau}+\nu_{\mu}^c$, depending on whether
$m_{\tau n}$ or $m_{\tau\mu}$ is large, respectively. In the former case,
$\theta_S\simeq m_{en}/m_{\tau n}$ while in the latter, $\theta_S\simeq
{m_{e\mu}/m_{\tau\mu}}$. The angle $\theta_n\simeq m_{\tau\mu}/m_{\tau n}$
(a) or $\theta_n\simeq m_{\tau n}/m_{\tau\mu}$ (b) determines the abundance of
$n$ during the nucleosynthesis. If it is less than $10^{-3}$, we expect
$N_{\nu}\leq 3.3$ and, if not, $N_{\nu}$ is close to 4.

The analysis for other choices of  $L$ can easily be performed along the same
lines and we do not present it here.
\section{The only good global symmetry is a broken global symmetry}
As we have seen up to now, in the limit of exact $L$ the neutrino spectrum
prevents oscillations in the light sector and so leaves the SNP unresolved.
On the other hand, the belief in exact global symmetries is becoming
increasingly less popular. This is certainly encouraged by the fact that the
virtual black holes and wormholes, while preserving local gauge invariance,
can destroy the
meaning of global quantum numbers. It is not unlikely then that there could be
higher dimensional operators cut off by the Planck scale which violate our
lepton number symmetry.  Barring the possibility of accidental cancellations
and assuming that the symmetry $G$ is not a part of a larger local gauge
symmetry, we expect this breaking to occur at the $d=5$ effective operator
level.

{\bf Neutrino mass}.
Without further ado then, we list the leading operators that could induce
corrections to the neutrino mass matrix \cite{HDO,GLR}:
\vspace*{.2truecm}
\EQ
(\nu_{i}^{T}C\tau_2\vec{\tau}\nu_i)\,\frac{H^{T}\tau_2\vec{\tau}H}
{M_{{\rm Pl}}},~~~
(n^TCn)\,\left[\frac{H^{\dagger}H}{M_{{\rm Pl}}}+
\frac{S^{2}}{M_{{\rm Pl}}}\right ]
\EN
\vglue .1truecm
\noindent
where $S^2$ stands for any bilinear combinations of the $S_{ab}$ fields, and
we list only the flavor-diagonal terms since their effect on $M_{\nu}$ is most
dramatic. Namely, they induce the mass splittings between the components of
Dirac and ZKM neutrinos and open up new channels for oscillations.

The split $\Delta m$ coming through the above operators when the relevant
fields get non-vanishing VEVs can be estimated as
\begin{equation} \Delta m\leq {M_W^2\over M_{\rm{Pl}}}\simeq 10^{-6}~{\rm eV},
\end{equation}
where the number $10^{-6}$ eV is probably an upper limit, since there could be
further dimensionless suppressions in (8) (certainly an order of magnitude
suppression should not come out as a surprise). We remind the reader that our
mass  scales are expected to lie close to the electroweak scale. An important
point is that the gravitational effects are expected to be flavor blind.
This implies that the mass splits in the light and heavy sectors should
be of the same order of magnitude. These splits $\Delta m$ being small, much
less than any Dirac mass term, leads to the prediction of two pseudo-Dirac
states with the mixing between the partners in each state being maximal
($\simeq 45^{\circ}$).

The squared mass differences in the light and heavy sectors will be
\EQ
\Delta m^{2}_{light} \simeq m_{\nu_e}\Delta m,~~~
\Delta m^{2}_{heavy} \simeq m_{17}\Delta m
\EN
where $\Delta m$ is given by eq. (8) and $m_{17}$ is the mass of heavy
neutrino. The experimental upper limit is $m_{\nu_e}<9.4$ eV \cite{LA} and,
as we shall see below from the discussion of the Majoron decays, there is a
lower limit $m_{\nu_e}>0.1$ eV implying $10^{-8}$ eV${}^2{}< \Delta
m^{2}_{light}{}<10^{-5}$ eV${}^2$. This range  allows for the neutrino
oscillations being naturally the solution of the SNP.

The oscillations in the heavy sector $\nu_{\mu}\rightarrow \nu_{\tau}(n^c)$
can be relevant for the recently reported deficiency of the atmospheric
$\nu_{\mu}$ \cite{ANP}. From eq. (9) it follows that for $m_{17}\sim 10$ keV
{}~$\Delta m^{2}_{heavy}$ can naturally be $\sim 10^{-2}-10^{-3}$ eV${}^2$
which
with the mixing angle being $45^{\circ}$ perfectly fits the required parameter
range \cite{ANP}.

The induced mass splittings of the pseudo-Dirac neutrinos open up new
channels of oscillations that can bring the sterile neutrino $n$ into the
equilibrium at the time of nucleosynthesis. Although the number of allowed
light species at that epoch is still debated \cite{Sarkar}, the frequently
quoted limit $N_{\nu}<3.4$ \cite{Chic}, if taken seriously, would imply
$\Delta m^{2}_{light}<5\times 10^{-9}$ eV${}^{2}$ if $n$ is a part of
$\nu_{light}$ and $\Delta m^{2}_{heavy}<8\times 10^{-6}$ eV${}^{2}$ when $n$
is a component of $\nu_{17}$ \cite{CL3}. From the limit $\Delta
m^{2}_{light}{}^>_\sim 10^{-8}$ eV${}^2$ it is clear that we are dangerously
close to the prediction of 4 light neutrino species in equilibrium, i.e.
$N_{\nu}$=4. However, due to the uncertainties in the estimation of the
gravitational effects, any conclusion would be premature; all we can say is
that $N_{\nu}$ could be lying anywhere between 3 and 4.

{\bf Majoron mass}.
As much as in the case of the neutrinos, we expect d=5 effective operators
explicitly violating lepton number in the scalar sector
\EQ
\frac{S}{M_{{\rm Pl}}}\left[S^4+S^2 H^{\dagger}H+(H^{\dagger}H)^2\right]
\EN
where we only give a typical example. Therefore the Majorons (in our case
flavons) acquire non-vanishing masses, i.e. become pseudo-Goldstone bosons.
Since we take $<S>\sim M_W$, we get an order of magnitude estimate
\EQ
m_F\simeq \left(\frac{M_{W}^3}{M_{{\rm Pl}}}\right)^{1/2}\simeq
1~{\rm keV}
\EN
The above is the typical value of the elements of the mass matrix of flavons
$F_{ab}$ which are expected to have large mixings with each other.

Since $m_F\ll m_{17}$, the decay rate of $\nu_{17}$ is almost unaffected
by the generated flavon masses. However, the issue now becomes whether
flavons themselves manage to decay fast enough to be in accord with
cosmological limits. Let us recall here the estimate of the $\nu_{17}$
lifetime due to the decay $\nu_{17}\rightarrow \nu_e+F$:
\EQ
\tau_{17}\simeq 16\pi \left(\theta_S\frac{m_{17}}{<S>}\right)^{-2}
m_{17}^{-1}\simeq 10^{-1}~{\rm s}
\EN
for $<S>\simeq M_W$, which is obviously cosmologically acceptable.
We should stress that $\nu_{17}$ is relativistic at the cosmological time
$t\simeq(0.1-1)$ sec and so the time dilation effect makes the actual
decay time in the comoving reference frame bigger than 1 sec. Therefore,
flavons appear only after the nucleosynthesis.\footnote{It is easy to see
that even if flavons were in equilibrium at $T\sim <S>\sim 100$ GeV for
whatever reason (e.g. due to the coupling to charged physical scalars
in certain models such as the one discussed in section 4), the weakness
of their interactions with neutrinos makes them decouple much before the
QCD phase transition. This means that they do not count at the time of
nucleosynthesis or, better to say, contribute a few percent of the usual
neutrino species.}

However, the cosmological problems related to the $\nu_{17}$ get now
replaced by the presence of massive flavons which are produced in the
$\nu_{17}$ decay with the concentration being equal to that of the active
neutrinos. The only possible mechanism to solve the problem of an
overabundance of massive flavons is their decay into light neutrinos
$\nu_{e}$. Recall that the coupling of Majorons to neutrinos is proportional
to the neutrino mass and this decay cannot take place for massless $\nu_e$.
This poses a serious problem for any Majoron-type models of the 17 keV
neutrino in which $\nu_e$ stays massless or very light (e.g. for
$L=L_{1,2,3}$, eq. (2) or in the absence of sterile neutrinos for
$L=L_{e}-L_{\mu}+L_{\tau}$). This question was recently raised by Grasso
{\em et al.} \cite{GLR} in the context of the BH picture. However, in our case
all we know is that $m_{\nu_{e}}<10$ eV \cite{LA} and so flavons are free to
decay into light neutrinos. As we show now, this provides a lower limit on the
$\nu_e$ mass\footnote{The general study of the cosmological implications of
massive Majorons is presently being done \cite{Mohapatra}.}.
It is easy to estimate the flavon lifetime due to the decay into two light
neutrinos:
\EQ
\tau_F\simeq 8\pi \left(\frac{m_{\nu_e}}{<S>}\right)^{-2}m_F^{-1}
\EN
which using eq. (11) becomes
\EQ
\tau_F \simeq 8\pi \frac{(<S>M_{{\rm Pl}})^{1/2}}{{m_{\nu_e}}^2}\simeq
10^{6} \left(\frac{{\rm eV}}{m_{\nu_e}}\right)^{2}~{\rm s}
\EN
It is clear from the above result that no useful limit on $\tau_F$ (i.e.
on $m_{\nu_e}$)
emerges from the requirement that the universe is not overclosed. A much
more serious constraint follows, however, from the galaxy formation
constraint. The recent COBE measurements of CMBR anisotropy \cite{COBE}
implies the small initial density fluctuations $\delta\rho/\rho\sim 10^{-5}$.
This, in turn, requires a sufficiently long matter-dominated epoch for the
linear growth of $\delta\rho/\rho$ to form the observed large-scale structure
of the universe. Therefore the decay products of the flavons have to be
redshifted sufficiently in order not to dominate the non-relativistic
matter (CDM) density at the time $t_{eq}$ of radiation dominated universe
turning into a matter dominant one in the standard picture. We, therefore,
demand
\EQ
m_Fn_{\nu}(t_{eq})\left(\frac{{\tau_F}}
{t_{eq}}\right)^{1/2}<\frac{1}{2}\rho_{cr}(t_{eq})
\EN
where $n_{\nu}(t_{eq})$ is neutrino number density at that time and
$\rho_{cr}(t_{eq})$ is the critical density at the same time. Using eq. (11)
one obtains the limit $\tau_F< 10^{7}$ s. This along with eq. (14) leads to
the promised lower limit on the electron neutrino mass
\EQ
m_{\nu_e}> 0.3~{\rm eV}
\EN
where due to uncertainties in the flavon masses and mixings, this limit should
be read as some number between 0.1 and 1 eV.

We should stress here that increasing the scale $<S>$ of the lepton
symmetry breaking only increases the lower limit on $m_{\nu_e}$ since
both $m_F$ and $\tau_F$ become larger. Moreover, at the scale $<S>\gg$
1 TeV flavons become heavier than $\nu_{17}$ and therefore $\nu_{17}$
itself cannot decay.

It is rather encouraging that the limit in (16) is not too far from the
laboratory upper limit on $m_{\nu_e}$. This provides even more
impetus for the direct experimental search of electron neutrino mass in
$\beta$ decays. It should be remembered that the almost Dirac nature of
$\nu_e$ in our work implies prediction of a negative result in the
experiments on neutrinoless double $\beta$ decay.

The cosmological implication of our result is also rather important.
Let us notice   that the concentration of light neutrinos
today is eight times that of a normal two-component neutrino. Recall that
before $\nu_{17}$ decay there are 4 light neutrino species, and this number
will not change with just the decay of $\nu_{17}$. However, the subsequent
decay of flavons adds yet another four species to the light neutrino
concentration of the present day universe\footnote{Each flavon decays into
a pair of light neutrinos, and in decay of $\nu_{17}$ two flavons are
produced and not one as it could appear at first sight. We are grateful
to J. Cline for this remark.}. So we can estimate the present-day light
neutrino concentration to be
\begin{equation}
n_{\nu}=8\times\frac{3}{11}n_{\gamma}\simeq 870~1/{\rm cm}^3
\end{equation}
where $n_{\gamma}\simeq 400/cm^3$ is the relic photon density of the
universe. Then from (16) we find that at least a fraction of the
present-day critical density of the universe is in the form of neutrinos, i.e.
hot dark matter. This observation may be rather important, since the standard
model of CDM with bias $b\simeq 2-3$ seems to be disfavored in view of recent
COBE measurements, if one takes the Harrison-Zeldovich $"$flat$"$ spectrum for
initial density fluctuations, which is motivated (up to some small
corrections) by inflation. In this case the value of the CMBR quadrupole
anisotropy is more than $2\sigma$ below the one that can be derived from the
COBE data \cite{COBE}. The latter is consistent with $b\simeq 1$, which seems
not to be in agreement with observed large scale structure, showing an excess
of a small scale power.

As it was shown in \cite{Shafi}, the partial (10--30\%) replacement of CDM by
HDM increases the large scale power and decreases the small scale one compared
to pure CDM case. So, such a CDM+HDM model, with $b\simeq 1$ and inflationary
(flat) spectrum can be made compatible with COBE data. Our situation, however,
differs somewhat from the one studied in ref. \cite{Shafi}, since there it was
assumed  only one light neutrino species with a mass $\sim 5-10$ eV, whereas
we end up with 8 times larger concentration and a mass $\sim 1$ eV. It would
follow that in our case one predicts the power spectrum to be more flat than
that in \cite{Shafi} for the same percentage of HDM. In any case this issue
deserves further consideration.

\section{The Model}
Although most of our analysis was to a large extent model independent, for
the sake of illustration we present a simple model based on $G=U(1)_e\times
U(1)_{\mu}\times U(1)_{\tau}\times U(1)_{n}$. The model is a straightforward
extension of that of BH \cite{BH}, which is based on the lepton flavor symmetry
in the Zee model \cite{Zee}. On top of $S_{ab}$ fields, one introduces a set
of $SU(2)_L$ singlet charged scalars $h_{ab}^{-}~(a\not= b)$ transforming as
$S_{ab}$ under $G$, which have the following couplings to leptons
\EQ
\Delta L_Y=f_{ij}l_i^TCi\tau_2 l_{j}h_{ij}^{*}+f_{in}(n^c)^TCe_i^c
h_{in}+h.c.
\EN
where $e_i^c$ are the charge conjugates of the right-handed leptons, singlets
under $SU(2)_L$. Note that here all the fermion fields are left-handed.

The transformation properties of $h_{ab}$ and $S_{ab}$ fields allow for an
important additional terms in the scalar potential
\begin{equation}
\Delta V=\lambda_{ab}(\phi_1^Ti\tau_2\phi_2)h_{ab}S_{ab}^{*}
+\lambda_{abcd}h_{ab}^{*}h_{cd}S_{ab}S_{cd}^{*}+h.c.
\end{equation}
where the two scalar doublets $\phi_1$ and $\phi_2$ are necessary because
of the antisymmetry of the $\phi\phi h$ coupling \cite{Zee}.

We illustrate here the case $L_+=L_e-L_{\mu}+L_{\tau}-L_{n^{c}}$, which
implies the only non-vanishing VEVs of the $S$ fields to be
\begin{equation}
<S_{e\mu}>\not= 0\not=<S_{\mu\tau}>,~~~<S_{\mu n}>\not= 0
\end{equation}
The leading radiatively induced neutrino masses are then (see Fig. 1)
$$m_{\mu\tau}\simeq\frac{1}{16\pi^2}(f_{\mu\tau}g_{\tau}
\lambda_{\mu\tau})m_{\tau}\frac{<S_{\mu\tau}><H>}{M^2} \nonumber $$
$$m_{e\mu}\simeq\frac{1}{16\pi^2}(f_{e\mu}g_{\mu}
\lambda_{e\mu})m_{\mu}\frac{<S_{e\mu}><H>}{M^2} \nonumber $$
$$m_{en}\simeq\frac{1}{16\pi^2}(f_{e\tau}f_{\tau n}
\lambda_{e\tau\tau n})m_{\tau}\frac{<S_{e\mu}><S_{\mu n}>}{M^2}
\nonumber$$
\EQ
m_{\tau n}\simeq\frac{1}{16\pi^2}(f_{\tau\mu}f_{\mu n}
\lambda_{\tau\mu\mu n})m_{\mu}\frac{<S_{\tau\mu}><S_{\mu n}>}{M^2}
\EN
\vglue .2truecm
\noindent
where $H$ is a linear combination of $\phi_1$ and $\phi_2$, chosen so as to
have $<H>\not=0$ and $g_i$ $(i=e, \mu, \tau)$ are the Yukawa couplings of the
orthogonal combination $\phi'$ with a vanishing VEV.  For simplicity we
assume all the scalar masses to be the same ($M$) \footnote{In any case, the
difference in the masses of the scalars can be reabsorbed in the unknown
coupling constants.}. Recall that the above pattern of VEVs can always be
achieved in the absence of additional symmetries.

Notice that for $M\simeq M_W, <S_{\mu\tau}>\simeq M_W$, with the
phenomenological limit $f_{\mu\tau}\leq 10^{-1}$, $m_{\mu\tau}$ is naturally
of the order 10 keV. Another way of phrasing this is that $M$ and $<S>$ must
be close to $M_W$ in order to reproduce the 17 keV neutrino. It is easy to
deduce an upper limit $<S>_\sim^<M_\sim^<1$ TeV, which implies that all the
new particles in the model can be detectable in the near future. This is the
most appealing  feature of the these type of models.

If $f_{e\mu}\sim f_{\mu\tau}$, $\lambda_{e\mu}\sim\lambda_{\mu\tau}$ one would
predict $\theta_S\sim m_{e\mu}/m_{\mu\tau} \sim m_{\mu}/m_{\tau} \simeq 0.1$.
Unfortunately, the predictions are obscured by the complete arbitrariness of
the couplings $g_i$ of the second doublet. Furthermore, the model as it stands
would not lead to the natural flavor conservation in the quark sector
\cite{DS}. The simplest way out is to have the doublets $\phi_{\alpha}$ to
couple to up and down quarks separately. This requires the existence of a
discrete symmetry $D$, such as
\EQ
\phi_u\rightarrow -\phi_u,~~~u_R\rightarrow -u_R,~~~S\rightarrow -S
\EN
and the rest of the fields invariant. Now obviously both $\phi_u$ and $\phi_d$
should have non-vanishing VEVs (due to an above symmetry, one cannot rotate
one  of the VEVs away). The spontaneously symmetry breaking of $D$ through
$<\phi_u>\not= 0$ leads at first glance to a catastrophic existence of the
domain walls. Fortunately, due to an anomaly, these walls can be shown to
decay away before dominating the energy density of the universe \cite{DW}.

With this $D$ symmetry one has
\EQ
g_i=\frac{m_i}{<\phi_d>}
\EN
and so to arrive at a correct value of $\theta_S$ one needs $\frac{f_{e\mu}}
{f_{\mu\tau}}\frac{\lambda_{e\mu}}{\lambda{\mu\tau}}\simeq 10$. Similar
adjustment is necessary to explain the hierarchy $m_{en}$, $m_{\tau n}\ll
m_{\mu\tau}$ which may be needed to comply with the cosmological limits on
the number of neutrino species. Again, there is enough freedom to accommodate
this requirement through the unknown $\lambda_{ijkl}$ couplings.
\section{Discussion}
In short, our model is a natural and straightforward extension of the BH
picture of flavons, i.e. Majorons associated with the lepton flavors. In the
limit of conserved $L$, the model is basically phenomenologically
indistinguishable from that of BH, except for possible cosmological role of
$n$. We do not repeat their analysis here, suffice it to mention their central
results:

(a) The "flavon" type models incorporate naturally 17 keV neutrino without
requiring any new mass scales.

(b) The most interesting prediction of BH which also holds here seems to be
the potentially observable $\tau\rightarrow eF$ ($F$ is a flavon) decay:
$BR(\tau\rightarrow eF)\simeq 10^{-4}$.

The principal motivation of our work was to attempt to shed some more light
on other central issues of neutrino physics, such as the problems of solar
and atmospheric neutrinos and the dark matter problem by adding a light
sterile neutrino. Of course, as long as the generalized ZKM lepton number
stays unbroken, one ends up with two 4-component neutrinos, one $\nu_{17}$
and another $\nu_{e}$ with mass ${}^<_\sim 10$ eV, and so no oscillations
relevant to the SNP and ANP are possible.

Once again we would like to stress the crucial nature of our gravitationally
induced breaking of $L$. Besides providing necessary mass splittings of the
order of $10^{-6}$ eV in both heavy ($\nu_{17}$) and light ($\nu_{e}$)
sectors, it also induces a substantial mass of flavons, of the order of 1 keV.
The requirement of sufficiently fast decay of flavons yields a lower bound on
the $\nu_{e}$ mass $m_{\nu_{e}}>(0.1-1)$ eV which, in turn, implies at least a
few percent of dark matter being hot.

Another important feature of our work is that squared mass difference in the
heavy neutrino sector is $\Delta m_{heavy}^2\sim 10^{-2}-10^{-3}$ eV${}^2$,
which together with maximal mixing is in the right range for the solution of
the ANP. This, however, can only work if $\nu_{17} \simeq\nu_{\tau}+
\nu_{\mu}^c$ since than $\nu_{\mu}\rightarrow \nu_{\tau}$ oscillations
can do the job \cite{ANP}. If the $\nu_{17}$ really exists, the ANP
can provide the
necessary insight into its structure. We would like to emphasize, though, that
its existence is by no means crucial for our work. It is true that without the
$\nu_{17}$ none of the other issues under consideration require the existence
of a light sterile state. It is only when gravitationally induced effects are
the source of the splittings of neutrino masses that $n$ is necessary for a
simultaneous solution of the SNP and ANP. We can even reverse the logic of our
analysis and say that the solution of the ANP in the context of Planck-scale
physics tends to suggest the existence of a neutrino in the 10 keV mass range.
Of course, its mixing angle with $\nu_{e}$ could easily be two orders of
magnitude smaller than $\theta_S$.

As was shown in section 3, $\Delta m_{light}^2$ lies in the range $10^{-8}-
10^{-5}$ eV${}^2$. This overlaps with the $\Delta m^2$ domain of the MSW
solution \cite{MSW} of the SNP. We should stress, however, that the MSW
effect is anyway irrelevant for the SNP in our scenario since the
mixing angle is practically equal to 45${}^{\circ}$. This means that we have
the short-wavelength vacuum oscillation $\nu_e\rightarrow n^c$ solution of the
SNP since $\Delta m^{2}_{light}\gg 10^{-10}$ eV${}^{2}$. Therefore one gets an
universal suppression factor $\simeq 1/2$ for all the solar neutrino
experiments. This is in a good agreement with the results of the Kamiokande
\cite{SNE2} SAGE \cite{SNE3} and GALLEX \cite{SNE4} but is at variance with
the Homestake data \cite{SNE1}. Further experiments are needed to clarify the
situation. The oscillation into a sterile state predicts suppressed neutrino
signals in the neutral-current channels in the forthcoming Sudbury Neutrino
Observatory \cite{SNO} experiment.

Our discussion up to now was almost exclusively devoted to the choice $L_+$ of
the conserved generalized ZKM symmetry. It is clear that the situation in the
case of $L_-$ is almost identical; some distinct features are listed in the
Table 1 (we should mention that all cases with $n$ in the heavy state are in
the potential conflict with the SN 1987A constraints \cite{SNB}, but we appeal
to new supernovae to resolve this issue). As far as the other choices, $L_1,
L_2$ and $L_3$ are concerned, they lead to one heavy and two massless
neutrinos (up to tiny gravitational effects inducing $\sim 10^{-6}-10^{-5}$ eV
masses for the latter) and so do not allow for the Majoron decays. Their
properties are still listed in the Table 1, since they naturally allow for the
so called "just so" oscillation solution of the SNP, with $\Delta m^2_{light}
\simeq 10^{-10}-10^{-11}$ eV$^2$ \cite{JS}. The natural way out of the Majoron
stability for these cases remains a challenge, since we do not wish to pursue
an unappealing possibility of fine-tuning the  flavon masses to be
sufficiently small.

Last but not least we wish to emphasize the relevance of the predicted
electron neutrino mass in the range 0.1-10 eV. To obtain what appears to be
a favored amount of about twenty per cent hot dark matter in the present-day
universe, $\nu_e$ mass should be approximately 1 eV which is in the reach of
a future direct observation.

\section*{Acknowledgements}
We would like to thank C. Burgess, J. Cline, A. Dolgov, R. Mohapatra,
S. Petcov, R. Schaefer, Q. Shafi, A. Smirnov and J. Valle for useful
discussions.

\newpage
{\small
\begin{tabular}{c|c|c|c|c|c}
\hline
\multicolumn{1}{c|}{$\begin{array}{c} Generalized~ ZKM \\ lepton~
number\end{array}$} &
\multicolumn{1}{c|}{$\begin{array}{c} Content~ of~ heavy \\ neutrino~
"\nu_{17}"\end{array}$} &
\multicolumn{1}{c|}{ANP} &
\multicolumn{1}{c|}{SNP} &
\multicolumn{1}{c|}{$N_{\nu}$} &
\multicolumn{1}{c}{HDM}\cr
\hline
\multicolumn{1}{c|}{ $L_+~(\theta_S\simeq\theta_{e\tau})$} &
\multicolumn{1}{c|}{$\begin{array}{c} \nu_{\tau}+\nu_{\mu}^c\\{}
\nu_{\tau}+n\end{array}$} &
\multicolumn{1}{c|}{$\begin{array}{c} \nu_{\mu}\leftrightarrow
\nu_{\tau}\\{}-\end{array}$} &
\multicolumn{1}{c|}{$\begin{array}{c} \nu_e\leftrightarrow n^c~(SW)\\{}
\nu_e\leftrightarrow\nu_{\mu}~(SW)\end{array}$} &
\multicolumn{1}{c|}{$\begin{array}{c} 3-4\\ 4 \end{array}$} &
\multicolumn{1}{c}{$\begin{array}{c} \nu_e\\ {}
\nu_e\end{array}$}\cr
\hline
\multicolumn{1}{c|}{ $L_-~(\theta_S\simeq\theta_{en^c})$} &
\multicolumn{1}{c|}{$\begin{array}{c} n^c+\nu_{\mu}^c\\{}
n^c+\nu_{\tau}^c\end{array}$} &
\multicolumn{1}{c|}{$\begin{array}{c} \nu_{\mu}\leftrightarrow n^c\\
{}-\end{array}$} &
\multicolumn{1}{c|}{$\begin{array}{c} \nu_e\leftrightarrow
\nu_{\tau}~(SW)\\{}
\nu_e\leftrightarrow\nu_{\mu}~(VA)\end{array}$} &
\multicolumn{1}{c|}{$\begin{array}{c} 4\\ 4 \end{array}$} &
\multicolumn{1}{c}{$\begin{array}{c} \nu_e\\ {}
\nu_e\end{array}$}\cr
\hline
\multicolumn{1}{c|}{ $~L_1\begin{array}{l} (\theta_S\simeq\theta_{e\tau})\\
(\theta_S\simeq\theta_{en^c})\end{array}$} &
\multicolumn{1}{c|}{$\begin{array}{c} \nu_{\tau}+\nu_{\mu}^c\\ n^c+
\nu_{\mu}^c\end{array}$} &
\multicolumn{1}{c|}{$\begin{array}{c} \nu_{\mu}\leftrightarrow
\nu_{\tau}\\ \nu_{\mu}\leftrightarrow n^c\end{array}$} &
\multicolumn{1}{c|}{$\begin{array}{c} \nu_e\leftrightarrow n^c~(JS)\\ {}
\nu_e\leftrightarrow\nu_{\tau}~(JS)\end{array}$} &
\multicolumn{1}{c|}{$\begin{array}{c} 3\\ 4\end{array}$} &
\multicolumn{1}{c}{$\begin{array}{c} ?\\ ?\end{array}$}\cr
\hline
\multicolumn{1}{c|}{ $L_2~~(\theta_S\simeq\theta_{en^c})$} &
\multicolumn{1}{c|}{$n^c+\nu_{\tau}^c$} &
\multicolumn{1}{c|}{$-$} &
\multicolumn{1}{c|}{$\nu_e\leftrightarrow\nu_{\mu}~(JS)$} &
\multicolumn{1}{c|}{4} &
\multicolumn{1}{c}{?}\cr
\hline
\multicolumn{1}{c|}{ $L_3~~(\theta_S\simeq\theta_{e\tau})~$} &
\multicolumn{1}{c|}{$\nu_{\tau}+n$} &
\multicolumn{1}{c|}{$-$} &
\multicolumn{1}{c|}{$\nu_e\leftrightarrow\nu_{\mu}~(JS)$} &
\multicolumn{1}{c|}{4} &
\multicolumn{1}{c}{?}\cr
\hline
\end{tabular}
\vskip 1cm

\noindent
Table 1. Summary of heavy neutrino composition and solutions for the SNP
and ANP for generalized lepton charges $L_+, L_-$ and $L_{1,2,3}$.
$SW$ and $JS$ stand for the solutions of the SNP through short wavelength
(averaged) vacuum oscillations ($\Delta m^2\simeq 10^{-8}-10^{-5}$ eV$^2$)
and "just so" oscillations ($\Delta m^2\simeq 10^{-10}-10^{-11}$eV$^2$)
respectively. Also shown are the effective number of neutrino species at the
time of nucleosynthesis $N_{\nu}$ and the composition of HDM. Question marks
indicate the problem with the decay of massive flavons explained in section 5.
\newpage
\centerline {\bf \large Figure caption}

\vglue 1truecm
\noindent
Fig. 1. One-loop diagrams which induce the neutrino mass terms $m_{ij}$
(a) and $m_{in}$ (b); $i,j$ take the values allowed by the $L$ symmetry.
$H$ and $\phi'$ are the linear combinations of $\phi_1$ and $\phi_2$
with non-vanishing and vanishing VEVs respectively.


\begin{thebibliography}{99}
\bibitem{SNE1} R. Davis, in Proc. XXI Int. Cosmic Ray Conf., ed. R.J.
Protheroe, vol. 12 (Univ. of Adelaide Press, Adelaide, Australia,
1990), p. 143.
\bibitem{SNE2} K.S. Hirata {\em et al.}, Phys. Rev. Lett. {\bf 65},
1297 (1990); Phys. Rev. D {\bf 44}, 2141 (1991).
\bibitem{SNE3} SAGE Collaboration, A.I. Abazov {\em et al.}, Phys. Rev.
Lett. {\bf 67}, 3332 (1991).
\bibitem{SNE4} GALLEX Collaboration,
P. Anselmann {\em et al.}, Phys. Lett. B {\bf 285}, 376, 390 (1992).
\bibitem{ANP} Kamiokande II collaboration, K.S. Hirata {\em et al.},
Phys. Lett. B {\bf 280}, 146 (1992); IMB Collaboration, D. Gasper
{\em et al.}, Phys. Rev. Lett. {\bf 66}, 2561 (1991); E.W. Beier
{\em et al.}, Phys. Lett. B {\bf 283}, 446 (1992).
\bibitem{COBE}G.F. Smoot {\em et al.}, COBE preprint, 1992
(Ap. J. Lett., {\em in press}).
\bibitem{Shafi} R.K. Schaefer, Q. Shafi, preprint BA-92-28(1992);
preprint IC/92/118, BA-92-45 (1992); A. van Dalen, R.K. Schaefer, preprint
BA-91-67 (1992).
\bibitem{PRO} J.J. Simpson, Phys. Rev. Lett. {\bf 54}, 1891 (1985);
Phys. Lett. {\bf 174}B, 113 (1986); J.J. Simpson, A. Hime,
Phys. Rev. D {\bf 39}, 1825 (1989); A. Hime, J.J. Simpson, Phys. Rev D
{\bf 39}, 1837 (1989); A. Hime, N.A. Jelley, Phys. Lett. B {\bf 257}, 441
(1991); B. Sur {\em et al}., Phys. Rev. Lett. {\bf 66}, 2444 (1991);
I. \v{Z}limen {\em et al}., Phys. Rev. Lett. {\bf 67}, 560 (1991).
\bibitem{CONTRA} T. Altzitzoglou {\em et al}., Phys. Rev. Lett. {\bf 55},
799 (1985); T. Ohi {\em et al}., Phys. Lett. B {\bf 160}, 322 (1985);
V.M. Datar {\em et al}., Nature (London), {\bf 318}, 547 (1985);
A. Apalikov {\em et al}., Pis'ma Zh. Eksp. Theor. Fiz. {\bf 42}, 233
(1985) [JETP Lett. {\bf 42}, 289 (1985)]; J. Markey, F. Boehm,
Phys. Rev. C {\bf 32}, 2215 (1985); D.W. Hetherington {\em et al}.,
Phys. Rev. C {\bf 36}, 1504 (1987); M.J.G. Borge {\em et al}.,
Phys. Scr. {\bf 34}, 591 (1986).
\bibitem{THEOR} See, e.g. S.L. Glashow, Phys. Lett. B {\bf 256}, 218
(1991); L. Bento, J.W.F. Valle, Phys. Lett. B {\bf 264}, 373 (1991);
K.S. Babu, R.N. Mohapatra, Phys. Rev. Lett. {\bf 67}, 1498 (1991);
K. Choi, A. Santamaria, Phys. Lett. B {\bf 267}, 504 (1991);
D. Choudhury, U. Sarkar, Phys. Lett. B {\bf 268}, 96 (1991);
G.K. Leontaris, C.E. Vayonakis, J.D. Vergados, Ioannina preprint IOA-
269, 1992; K.S. Babu R.N. Mohapatra, I.Z. Rothstein, Phys. Rev. D {\bf 45},
R5 (1992). In the last paper a comprehensive list of literature on the
subject can be found.
\bibitem{X} D.O. Caldwell, P. Langacker, Phys. Rev. D {\bf 44}, 823
(1991); A.Yu. Smirnov, talk at the EPS Conference, CERN, July 1991
(unpublished); A.Yu. Smirnov, J.W.F. Valle, preprint FTUV/91-38,
1991; J. Peltoniemi, A.Yu. Smirnov, J.W.F. Valle, preprint FTUV/92-6,
1992; K.S. Babu, R.N. Mohapatra, I.Z. Rothstein, Phys. Rev D {\bf 45},
R5 (1992); K.S. Babu, R.N. Mohapatra, preprint UMDHEP 92-150, 1992;
E. Ma, Phys. Rev. Lett. {\bf 68}, 1981 (1992); C.P Burgess, J.M. Cline,
M.A. Luty, Phys. Rev. D {\bf46}, 364 (1992).
\bibitem{ZKM} Ya.B. Zeldovich, Dokl. Akad. Nauk SSSR {\bf 86},
505 (1952); E.J. Konopinski, H. Mahmoud, Phys. Rev. {\bf 92}, 1045 (1953).
\bibitem{V} M.J. Dugan, G. Gelmini, H. Georgi, L.J. Hall, Phys. Rev.
Lett. {\bf 54}, 2302 (1985); J.W.F. Valle, Phys. Lett. B {\bf 159},
49 (1985). This lepton charge was first introduced in another context by
S.T. Petcov, Phys. Lett. {\bf 110}B, 245 (1982).
\bibitem{LEP} ALEPH Collaboration, D. Decamp {\em et al}., Phys. Lett.
B {\bf 235}, 399 (1990); OPAL Collaboration, M.Z. Akrawy {\em et al}.,
Phys. Lett. B {\bf 240}, 497 (1990); DELPHI Collaboration, P. Abreu
{\em et al}., Phys. Lett. B {\bf 241}, 435 (1990); L3 Collaboration,
B. Adeva {\em et al}., Phys. Lett. B {\bf 249}, 341 (1990).
\bibitem{BH} R. Barbieri, L. Hall, Nucl. Phys. {\bf B364}, 27 (1991);
Z. Berezhiani, (unpublished).
\bibitem{HDO} R. Barbieri, J. Ellis, M.K. Gaillard, Phys. Lett.
{\bf 90}B, 249 (1980); E. Akhmedov, Z. Berezhiani, G. Senjanovi\'{c},
preprint IC/92/79, SISSA-83/92/EP, LMU-04/92.
\bibitem{GLR} D. Grasso, M. Lusignoli, M. Roncadelli, Rome preprint
n.868, FNT/T 92/07, 1992.
\bibitem{SNB} G. Raffelt, D. Seckel, Phys. Rev. Lett. {\bf 60}, 1793
(1988); K.J.F. Gaemers, R. Gandhi, J.M. Lattimer, Phys. Rev D {\bf 40},
309 (1989); J.A. Grifols, E. Mass\'{o}, Phys. Lett. B {\bf 242}, 77 (1990);
R. Gandhi, A. Burrows, Phys. Lett. B {\bf 246}, 149 (1990); B {\bf 261},
519(E) (1991); M.S. Turner, Phys. Rev. D {\bf 45}, 1066 (1992);
S. Dodelson, J.A. Frieman, M.S. Turner, Phys. Rev. Lett. {\bf 68}, 2572
(1992); A. Burrows, R. Gandhi, M.S. Turner, Phys. Rev. Lett. {\bf 68},
3834 (1992).
\bibitem{SpM} See, e.g. A.Yu. Smirnov and J.W.F. Valle, ref. [4]. The
phenomenology of the system with the conserved lepton charge $L_{+}$
(with $n$ being an active neutrino of the fourth generation) was analysed in
a model-independent way by M. Lusignoli, Phys. Lett. {\bf 168}B, 307 (1985).
\bibitem{CL1} A.D. Dolgov, Yad. Fiz. {\bf 33}, 1309 (1981) [Sov. J. Nucl.
Phys. {\bf 33}, 700 (1981)]; M.Yu. Khlopov, S.T. Petcov, Phys. Lett. B
{\bf 99}, 117 (1981); B {\bf 100}, 520 (1981) (E);  A.D. Dolgov,
D.P. Kirrilova, Intern. J. Mod. Phys. A {\bf 3}, 267 (1988).
\bibitem{CL2} R. Barbieri, A. Dolgov, Phys. Lett. B {\bf 237}, 440 (1990);
Nucl. Phys. {\bf B349}, 743 (1991); K. Kainulainen, Phys. Lett. B {\bf 244},
191 (1990); K. Enqvist, K. Kainulainen, J. Maalampi, Nucl. Phys. {\bf B349},
754 (1990).
\bibitem{CL3} K. Enqvist, K. Kainulainen, M. Thomson, Phys. Rev. Lett.
{\bf 68}, 744 (1992); Nucl. Phys. {\bf B373}, 498 (1992).
\bibitem{Chic} K.A. Olive, D.N. Schramm, G. Steigman, T.P. Walker, Phys. Lett.
B {\bf 236}, 454 (1990).
\bibitem{LA} J.F.Wilkerson et al, in Proceedings of the 14th International
Conference on Neutrino Physics and Astrophysics, CERN, Geneva, June 10-15,
1990 [Nucl.Phys.B Proc.Suppl. {\bf 19}, 215, 1991].
\bibitem{Sarkar} A. Hime, R.J.N. Phillips, G.G. Ross, S. Sarkar, Phys. Lett.
B {\bf 260}, 381 (1991).
\bibitem{Mohapatra}  E.Kh. Akhmedov, Z.G. Berezhiani, R.N. Mohapatra and
G. Senjanovi\'c, in preparation.
\bibitem{Zee} A. Zee, Phys. Lett. {\bf 93}B, 389 (1980); {\bf 161}B, 141
(1985).
\bibitem{DS} S.L. Glashow, S. Weinberg, Phys. Rev. D {\bf 15}, 1958 (1977);
E.A. Paschos, Phys. Rev. D {\bf 15}, 1966 (1977).
\bibitem{DW} J. Preskill, S.T. Trivedi, F. Wilczek, M.B. Wise, Nucl. Phys.
{\bf B363}, 207 (1991).
\bibitem{MSW} S.P. Mikheyev, A.Yu. Smirnov, Yad Fiz {\bf 42}, 1441 (1985)
[Sov. J. Nucl. Phys. {\bf 42}, 913 (1985)]; L. Wolfenstein, Phys. Rev. D
{\bf 17}, 2369 (1978).
\bibitem{SNO} SNO Collaboration, G. Aardsma {\em et al.}, Phys. Lett. B
{\bf 194}, 321 (1987).
\bibitem{JS} S.M. Bilenky, B. Pontecorvo, Phys. Rep. {\bf 41}, 225 (1978);
V. Barger, R.J.N. Phillips, K. Whisnant, Phys. Rev. D {\bf 24}, 538 (1981);
Phys. Rev. D {\bf 43}, 1110 (1991); S.L. Glashow, L.M. Krauss, Phys.Lett. B
{\bf 190}, 199 (1987); A. Acker, S. Pakvasa, J. Pantaleone, Phys. Rev. D
{\bf 43}, 1754 (1991); P.I. Krastev, S.T. Petcov, preprint CERN-TH. 6401/92
(1992).
\end{thebibliography}
\end{document}